\documentclass[10pt,journal,compsoc]{IEEEtran}

\ifCLASSOPTIONcompsoc
  \usepackage[nocompress]{cite}
\else
  \usepackage{cite}
\fi

\usepackage{cite}
\usepackage{amsmath,amssymb,amsfonts}
\usepackage{algorithmic}
\usepackage{graphicx}
\usepackage{textcomp}
\usepackage{xcolor}
\usepackage{subfigure}
\usepackage{graphicx}
\usepackage[ruled,vlined]{algorithm2e}
\usepackage{hyperref}
\usepackage{threeparttable}

\begin{document}

\title{Wound Segmentation with Dynamic Illumination Correction and Dual-view Semantic Fusion}
\author{Honghui~Liu,
        Changjian~Wang*\thanks{*Corresponding authors},
        Kele~Xu*, 
        Fangzhao~Li,
        Ming~Feng,
        Yuxing~Peng,
        Hongjun~He
\IEEEcompsocitemizethanks{\IEEEcompsocthanksitem Honghui~Liu, Changjian~Wang, Kele~Xu, Fangzhao~Li,  Ming Feng, Yuxing~Peng, Hongjun~He are with National University of Defense Technology, Changsha, China.}
\IEEEcompsocitemizethanks{\IEEEcompsocthanksitem Ming~Feng is with Tongji University, Shanghai, China.}
\thanks{Manuscript received April XX, 2022; revised April XX, 2022.}}

\markboth{Journal of \LaTeX\ Class Files,~Vol.~14, No.~8, August~2015}%
{Shell \MakeLowercase{\textit{et al.}}: Bare Demo of IEEEtran.cls for Computer Society Journals}

\IEEEtitleabstractindextext{%
\begin{abstract}
Wound image segmentation is a critical component for the clinical diagnosis and in-time treatment of wounds. Recently, deep learning has become the mainstream methodology for wound image segmentation. However, the pre-processing of the wound image, such as the illumination correction, is required before the training phase as the performance can be greatly improved. The correction procedure and the training of deep models are independent of each other, which leads to sub-optimal segmentation performance as the fixed illumination correction may not be suitable for all images. To address aforementioned issues, an end-to-end dual-view segmentation approach was proposed in this paper, by incorporating a learn-able illumination correction module into the deep segmentation models. The parameters of the module can be learned and updated during the training stage automatically, while the dual-view fusion can fully employ the features from both the raw images and the enhanced ones. To demonstrate the effectiveness and robustness of the proposed framework, the extensive experiments are conducted on the benchmark datasets. The encouraging results suggest that our framework can significantly improve the segmentation performance, compared to the state-of-the-art methods.

\end{abstract}

}

\maketitle

\IEEEdisplaynontitleabstractindextext

\IEEEpeerreviewmaketitle

\IEEEraisesectionheading{\section{Introduction}\label{sec:introduction}}

\IEEEPARstart{W}{ound} image segmentation is an important research topic in the medical image processing field \cite{minaee2021image,cheng2021boundary}, which is valuable for the clinical diagnosis and treatment. Delayed treatment inevitably affects the physical and mental health of the patients, and it may take years to heal and sometimes the wound would re-occur. Effective evaluation of the wound in the images can be greatly helpful to alleviate aforementioned issues \cite{ghosh2019understanding,kirillov2020pointrend}.

Since the renaissance of deep neural networks that began in 2006, deep learning has been increasingly used for wound image segmentation \cite{zhou2019unet,isensee2021nnu}, and it has achieved excellent performance on multiple benchmarks. Wang et al.\cite{wang2015a} proposed an encoder-decoder architecture, which employed the deep convolutional neural network (CNN) for the wound segmentation. Leveraging the versatile representation learning ability, the network could identify different kinds of wounds, and provide higher accuracy than the traditional segmentation approaches. However, these methods had an apparent shortcoming as they did not consider the complexity of wound image background, and it was difficult to distinguish the object on the background which is similar to the wounds. To address the aforementioned issue, Li et al.\cite{li2018composite} used Cr channel in YCbCr color space and combined with deep learning network to identify skin area to limit results of segmentation in skin area. This method could partly eliminate the influence of complex backgrounds and improve the accuracy of segmentation.

\begin{figure*}[h] 
		\centering 
		\subfigure[]{
			\label{fig1(a)}
			\includegraphics[width=1.15in]{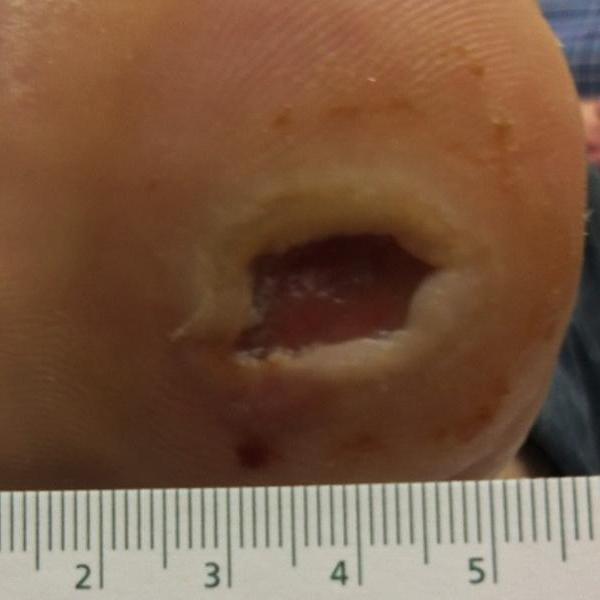}} 
		\hspace{0.2in}
		\subfigure[]{
			\label{fig1(b)}
			\includegraphics[width=1.15in]{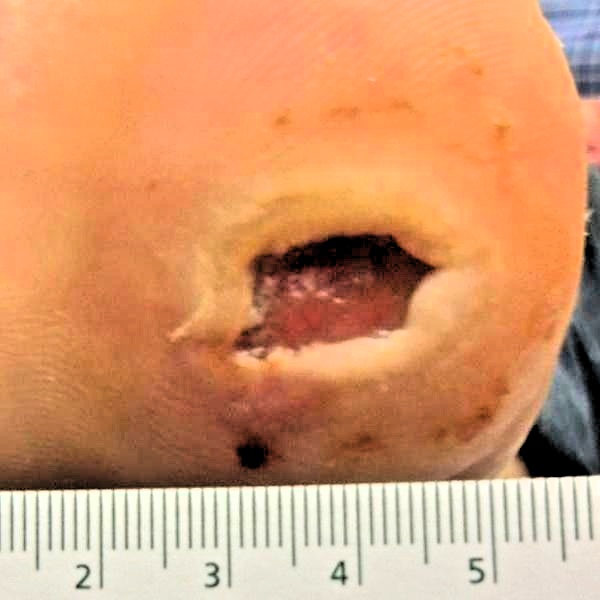}}
		\hspace{0.2in}
		\subfigure[]{
			\label{fig1(c)}
			\includegraphics[width=1.15in]{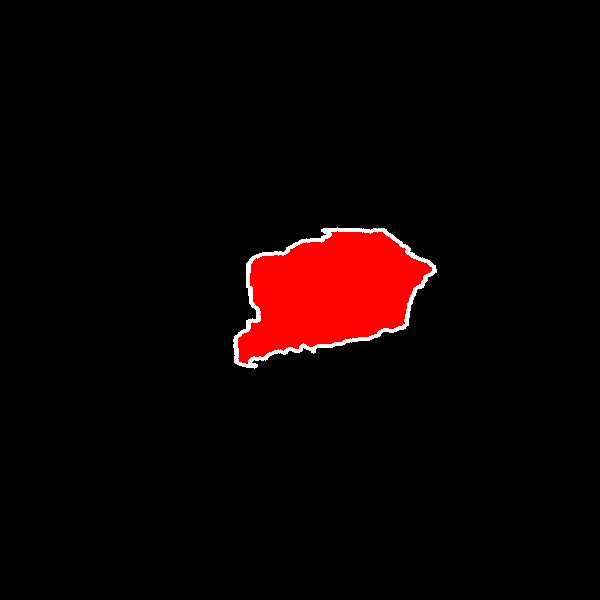}}
		\hspace{0.2in}
		\subfigure[]{
			\label{fig1(d)}
			\includegraphics[width=1.15in]{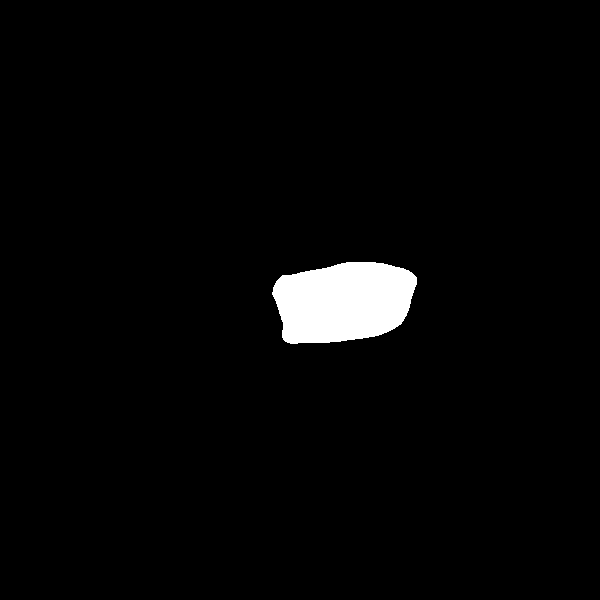}}
		\hspace{0.2in}
		\subfigure[]{
			\label{fig1(e)}
			\includegraphics[width=1.15in]{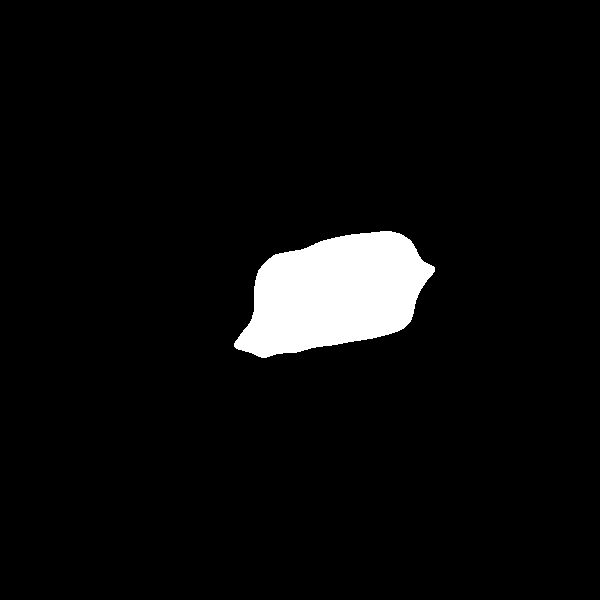}}
		\caption{A running example to demonstrate the effectiveness of the illumination correction module. (a) Original image; (b) Illumination Corrected image; (c) Ground Truth for the segmentation; (d) Segmentation result obtained without illumination correction module; (e) Segmentation result obtained with our illumination correction module.} 
		\label{fig1}
	\end{figure*}
	
Despite the sustainable efforts being made, many of the previous methods did not consider the medical imaging conditions as the imaging conditions may vary in practical settings. For example, the illumination inhomogeneity might appear in the wound images ~\cite{xie2020sesv}, which could decrease the performance of segmentation ~\cite{wang2018interactive,zhao2021ant,jang2019interactive,li2020image}. To address this problem, Lu et al.~\cite{lu2017wound} utilized a traditional algorithm to correct the illumination inhomogeneity, followed by the training of the network of wound segmentation. This method improves the accuracy of those images whose illumination is inhomogeneous. As a running example, Figure~\ref{fig1} illustrates that the segmentation performance can be improved after correcting the illumination. For many of previous attempts with illustration correction, the illumination and training phase are two independent procedures, thus the correction is conducted using the static manner, which aims at improving the segmentation pheromone in the images with inhomogeneous illumination. The images with good illumination condition may have color distortion, so that the accuracy would decrease.

Unlike previous static pre-processing mechanisms, this paper proposes to incorporate a dynamic illumination correction module into the deep segmentation network~\cite{losel2020introducing,lalonde2021capsules}. This module uses convolution operation to filter intensity channels of image, thus the parameters can be learned during the training phase~\cite{ye2019cross}. To improve the performance further, we proposed a dual-view segmentation structure network, to retain the features of original image and corrected image simultaneously and then combine the multi-scale features in the expansive path. Combined with these two improvements, the method proposed in this paper exhibits superior performance on the benchmarks. 


The main contributions of this paper can be summarized as follows: (1) Firstly, we proposed an end-to-end wound image segmentation method, by incorporating a dynamic illumination correction module. Specifically, this network converts the multi-scale Gaussian filtering into the learn-able convolution layers of the neural network. Combining the correction and segmentation network, the illumination module and segmentation can be optimized simultaneously.\\
(2) Furthermore, to retain the salient features both from the original image and the corrected image, a dual-view segmentation fusion strategy is explored in this paper. Both the representative features can be extracted in the contraction path and further combine in the expansive path.\\
(3) To demonstrate the performance of the proposed solution, we conducted extensive experiments on the benchmark datasets. The experimental results suggested that our proposed method is capable of archiving state-of-the-art performance under different experimental settings.

The remainder of this paper is organized as follows. In Section 2, the  related works are discussed while Section 3 formally details our methodology. We conduct extensive experiments to evaluate the effectiveness and robustness of our proposed method in Section 4. In the end, the concluding remarks are given in Section 5.

\section{Related Work}

\subsection{Wound Segmentation}
The segmentation of wound images plays a crucial role in the clinical diagnosis and treatment~\cite{zhao2019data,sinha2020multi,gu2019net,chen2018drinet}. Many researches have been carried out on this problem, and some satisfied performance has been obtained. The current methods for this problem are mainly divided into two categories: the traditional methods and the deep learning methods. The previous methods of wound segmentation mainly use traditional semantic segmentation algorithms \cite{10.1007/11559573_123,song2012automated,feng2020cpfnet,hu2019topology,wang2018interactive} which includes threshold-based method, active contour model, watershed method and so on. The emergence of deep convolutional neural networks makes deep learning methods more suitable for the medical image segmentation field \cite{krizhevsky2012imagenet,simonyan2014very}. FCN~\cite{7478072} explored deep learning for semantic segmentation. Presently, deep learning methods have become the mainstream of wound segmentation \cite{chen2014semantic,ronneberger2015u-net:}. This type of method significantly improves the accuracy of the wound segmentation algorithm, and has good generalization ability.

Wang et al~.\cite{wang2015a} designed a multi-layer CNN leveraging the encoder-decoder structure. The network could identify different types of wounds and introduce deep learning methods into the field of wound segmentation. The accuracy is higher than that of traditional methods, but it still does not meet clinical standards and there is still huge space to improve~\cite{zhou2019unet,kirillov2020pointrend,ghosh2019understanding,chen2018encoder}. The traditional semantic segmentation network cannot be applied to wound images directly. One of the important reasons is that wound image does not have a large-scale dataset to achieve excellent training results. Moreover, some images in the datasets have poor quality, and many wound images do not have a good shooting environment and professional photographic equipment. Therefore, a certain degree of pre-treatment is indispensable for such problems. Li et al.~\cite{li2018a} used the CR channel in YCbCr pattern to distinguish the skin area in the image, and combined it with a deep learning network. It enables the wound segmentation limited in the skin area, eliminates the influence of part of the complex background, and improves the recognition accuracy. Lu et al.\cite{lu2017wound} noticed that uneven illumination has a profound impact on the final segmentation results which uses traditional image algorithms to correct the light intensity and color, and then performs deep learning training. The algorithm architecture of pre-processing and training also has practical effects \cite{minaee2021image,li2018composite,ibtehaz2020multiresunet}.

\subsection{Illumination Inhomogeneity}
Despite the sustainable efforts been made, many of the previous methods did not consider the configuration setting during the medical imaging as the imaging condition is not unified. For example, the illumination inhomogeneity may appear in the images, which unavoidable decrease the performance of segmentation. There are many algorithms to deal with uneven illumination, such as histogram equalization (HE), gamma correction and so on. Presently, the most effective Illumination correction method is based on Retinex theory, which is inspired \cite{land1971lightness} based on human perception of the scene. The main idea is to divide an image into two parts, one is illumination, the other is the reflection property of the object itself. It can be expressed as:

\begin{equation}
	I_{i}(x,y)=S_{i}(x,y) * R_{i}(x,y) \label{eq1}
\end{equation}
where \emph{I$_i$} is the input image on the \emph{i-th} color channel and \emph{S$_i$} is the illumination and \emph{R$_i$} is the scene reflectance. From the theoretical perspective, the main purpose of the variants is to remove the illumination component and balance the illumination conditions of the image. Sustainable efforts have been made, and the state-of-the-art algorithm of \emph{msrcp}~\cite{jobson1997multiscale,rahman1996multi} is built on the channel of {intensity} to reduce the number of convolution layers. The algorithm of \emph{simple color balance} is used to simplify the process of transforming the number field into the real field. Despite the simplicity and the effectiveness of the illumination correction, how to incorporate the illumination correction module into a learn-able segmentation network is still under-explored in previous studies.

\section{Methodology}
As aforementioned, the pre-processing part is critical to improve the performance of the wound image segmentation for different imaging conditions. Despite the efforts have been made, two challenges for current wound segmentation methods are still need to be addressed. Firstly: the pre-processing algorithm is fixed and its parameters cannot be adjusted according to the training process, thus leads to sub-optimal segmentation performance. Secondly: the traditional pre-processing may decrease the segmentation accuracy for well-illuminated images, thereby affecting the overall accuracy. To address these two challenges, we propose a novel framework for the wound segmentation task. Specifically, compared to standard segmentation architectures, two novel strategies are proposed: dynamic illumination correction and dual-view semantic fusion. We will explain the strategies in more details subsequently.
	
\subsection{Dynamic illumination correction}
\begin{figure}[]
	\centering
	\includegraphics[width=9.0cm]{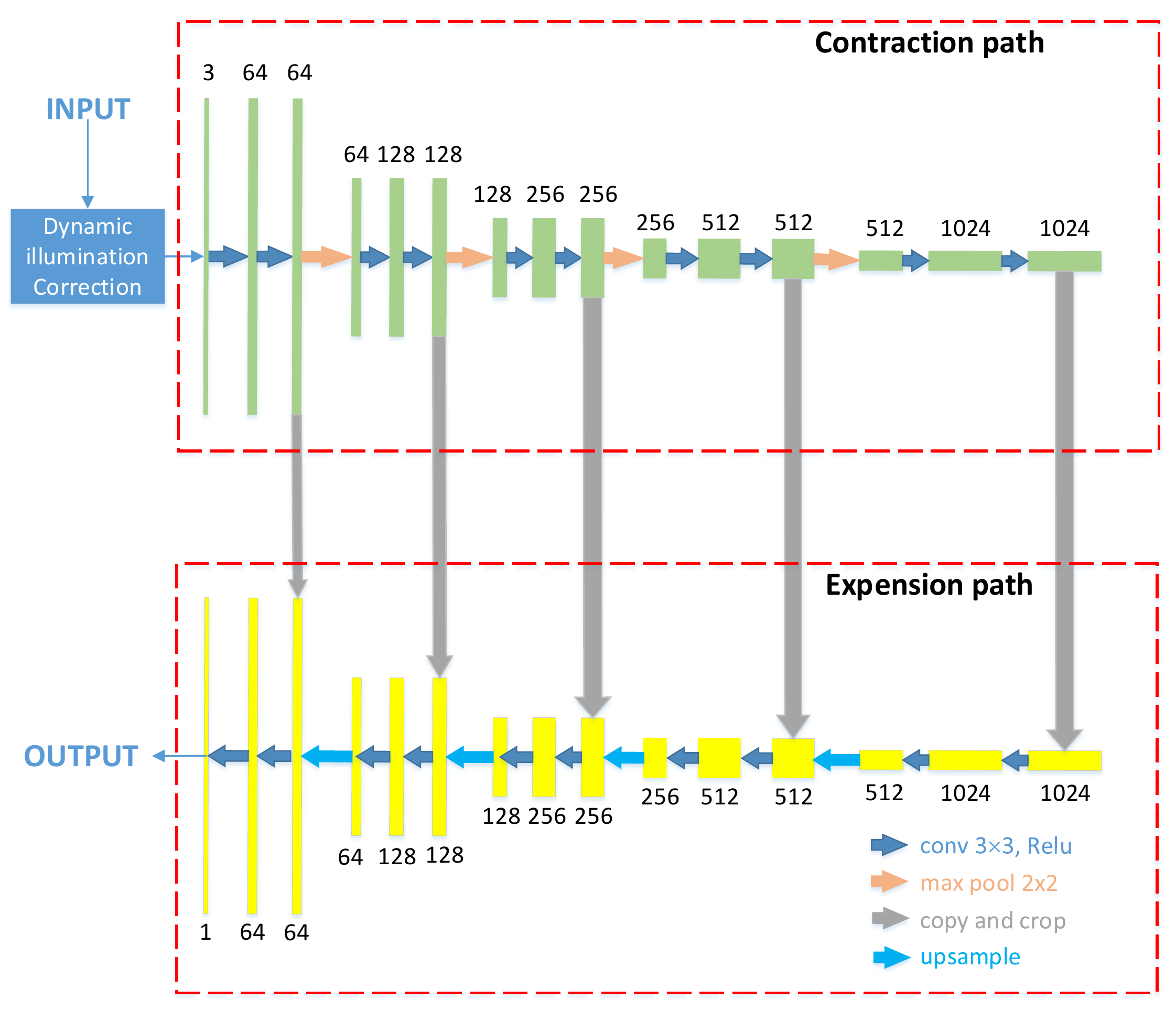}
	\caption{DICU, the flowchart of the Dynamic illumination correction module.}
	\label{fig2}
\end{figure}

In our framework, an end-to-end learn-able illumination correction module was employed for the segmentation task. Here, we denote the network as \textbf{DICU}, which means dynamic illumination correction in U-Net, as shown in Fig. \ref{fig2}. The network can conduct the pre-processing using the dynamic manner, whose key idea is to correct the illumination using a learn-able pre-processing module. 

In our dynamic pre-processing module, we model the Gaussian filtering using the convolution operations with variable parameters. As is well-known, convolution layer is one of the most important essential elements for CNNs. In this way, the pre-processing can be easily integrated into convolutional neural network. Thus, the pre-processing module algorithm\ref{alg.1} can be simple, fast, and has multiple functions besides illumination correction, such as increasing the contrast and sharpening the border.

\begin{equation}
	F(x,y)=C \times exp[-(x^2+y^2)/2\sigma^2] \label{eq2}
\end{equation}
where $\sigma$ is the scale of Gaussian Filter,and c is the parameter of normalization.
\begin{equation}
	\int_{}^{} F(x,y) dxdy = 1 \label{eq3}
\end{equation}

For a typical RGB image, the intensity channel is required firstly, which calculates the average value of each channel in each pixel. Then, three convolutions operations are used with different sizes filter the intensity channel and the size of the convolutions ranges from are 12, 80 and 250. Convolution from multiple scales can not only take into edge details account, but also solve the problem of dynamic range compression. As a result, the multi-scale relative illumination intensity of the image can be obtained. After that, the relative light intensity needs to be removed to get a light-corrected image. In order to simplify the calculation, the channel value is converted from the real number domain to the logarithmic domain. By this way, the matrix division operation can be converted to subtraction, which greatly simplifies the calculation amount. In this way, the intensity channel after illumination correction is obtained. Then use the Simplest Color Balance algorithm \ref{alg.2} to cut out the percentage of the pixel values at both sides of the value range, which can be converted into a real number domain by linear operation and fill the interval of [0, 255].

\begin{figure*}[h]
	\centering
	\includegraphics[width=17cm]{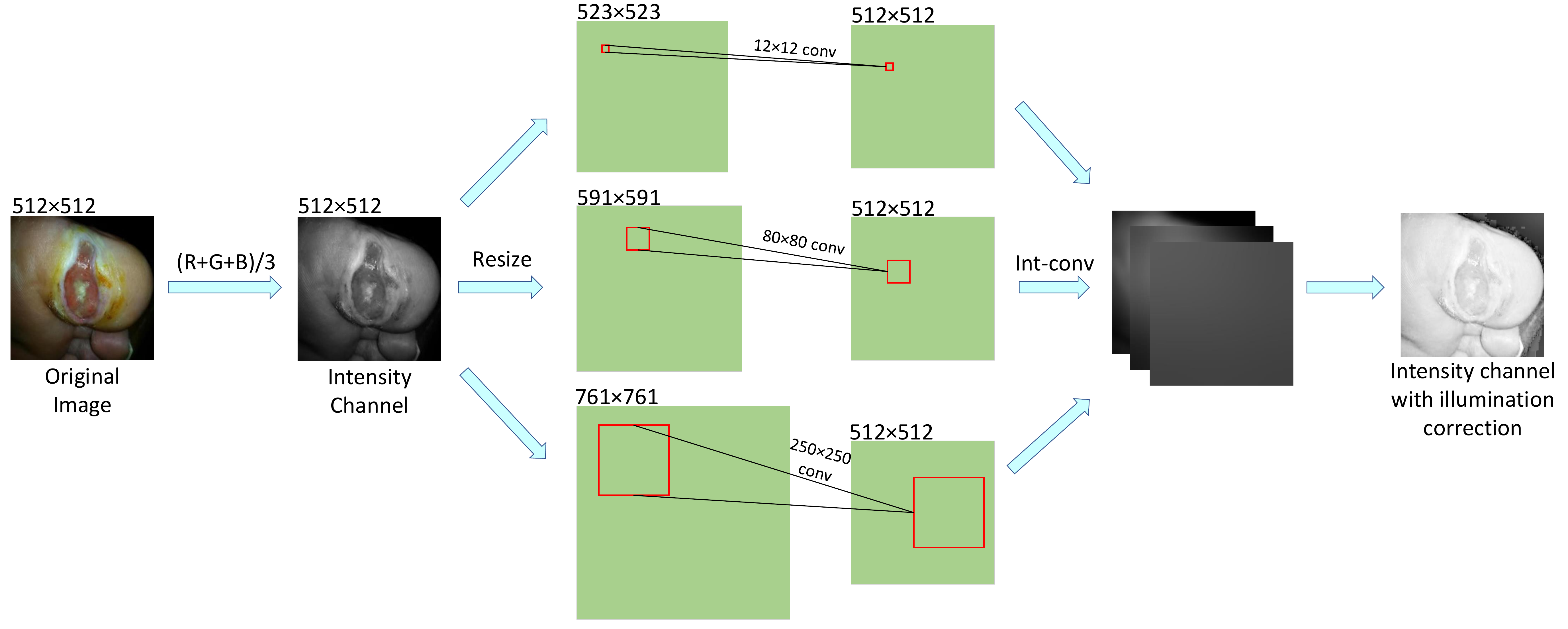}
	\caption{Large convolution kernels using the resize operation.}
	\label{fig3}
\end{figure*}

\begin{algorithm}[!t]
	\caption{Illumination Correction}
	\hspace*{0.00in} {\bf Data:} 
	\emph{I}:input color image;  $\sigma_1$,$\sigma_2$,$\sigma_3$:the scales; s$_1$,s$_2$ the percentage of clipping pixels on each side\\
	\hspace*{0.00in} {\bf Result:} 
	Illumination correction image;\\
	$Int = (I_R+I_G+I_B)/3$\\
	\ForEach{$\sigma_i$}{
		$Diff_i=log(Int)-log(Int*Conv_{\sigma_i})$
	}
	$MSR=\sum_{i=1}^{3}Diff_i$\\
	$Int_1=SimplestColorBalance(MSR,s_1,s_2)$\\
	\ForEach{pixel i}{
		$B=max(I_R[i],I_G[i],I_B[i])$\\
		$A=min(\frac{255}{B},\frac{Int_1[i]}{Int[i]})$\\
		$Illu\_Crec_R[i]=A*I_R[i]$\\
		$Illu\_Crec_G[i]=A*I_G[i]$\\
		$Illu\_Crec_B[i]=A*I_B[i]$
	}
    \label{alg.1}
\end{algorithm}
In this way, the intensity channel can be obtained by removing the light component and named it Int1. To use the ratio of Int1 to the original intensity channel Int to scale the pixel value of the original image in proportion to get the result. In the illumination correction module, three large convolution kernels are applied, and the scales of which are 12, 80 and 250. The parameters of these three convolution kernels are constantly adjusted during the training process, thereby realizing the dynamic adjustment algorithm, so that the module can better adapt to the data set. It is expected that after the convolution operation, the feature map still maintains its original size. For a large-size convolution kernel, padding will make the edge of the feature map have a lot of zeros, which will greatly affect the accuracy of segmentation. In order to solve this problem, the image was preprocessed for different sizes of convolution kernels. Using the Resize function, the image is enlarged to a certain ratio, so that after three large convolution kernels, the feature map can be restored to its original size, (Figure \ref{fig3}).  \\

\begin{algorithm}[!t]
	\caption{Simplest Color Balance}
	\hspace*{0.02in} {\bf Data:}
	I: Multi-Scale Retinex Channel; s$_1$, s$_2$ are the percentage of clipping pixels on each side\\
	\hspace*{0.02in} {\bf Result:}
	Int1\\
	$N=H*W$\\
	$sort_I=sort(I)$\\
	$Vmin=sort_I[N*s1]$\\
	$Vmax=sort_I[N*(1-s2)]$\\
	\ForEach{pixel i}{
		\If{i$>$Vmax}{
			$i=Vmax$
		}
		\If{i$<$Vmin}{
			$i=Vmin$
		}
		$i=(i-Vmin)*255/(Vmax-Vmin)$
	}
	\label{alg.2}
\end{algorithm}

\subsection{Dual-view semantic fusion}
Another issue with current wound segmentation algorithms is that: the preprocessing algorithm aimed at illumination correction cannot be directed used for all images. There are still many images in the dataset of good quality and in a good illumination condition. The preprocessing algorithm for illumination correction is very likely to bring out the overexposure and color imbalance of those images, resulting in a decrease of the segmentation performance. Here, a dual-view network was designed, which is called \textbf{DVSFN} in this paper, which means dual-view semantic fusion network.

\begin{figure}[htbp]
	\centering
	\includegraphics[width=9cm]{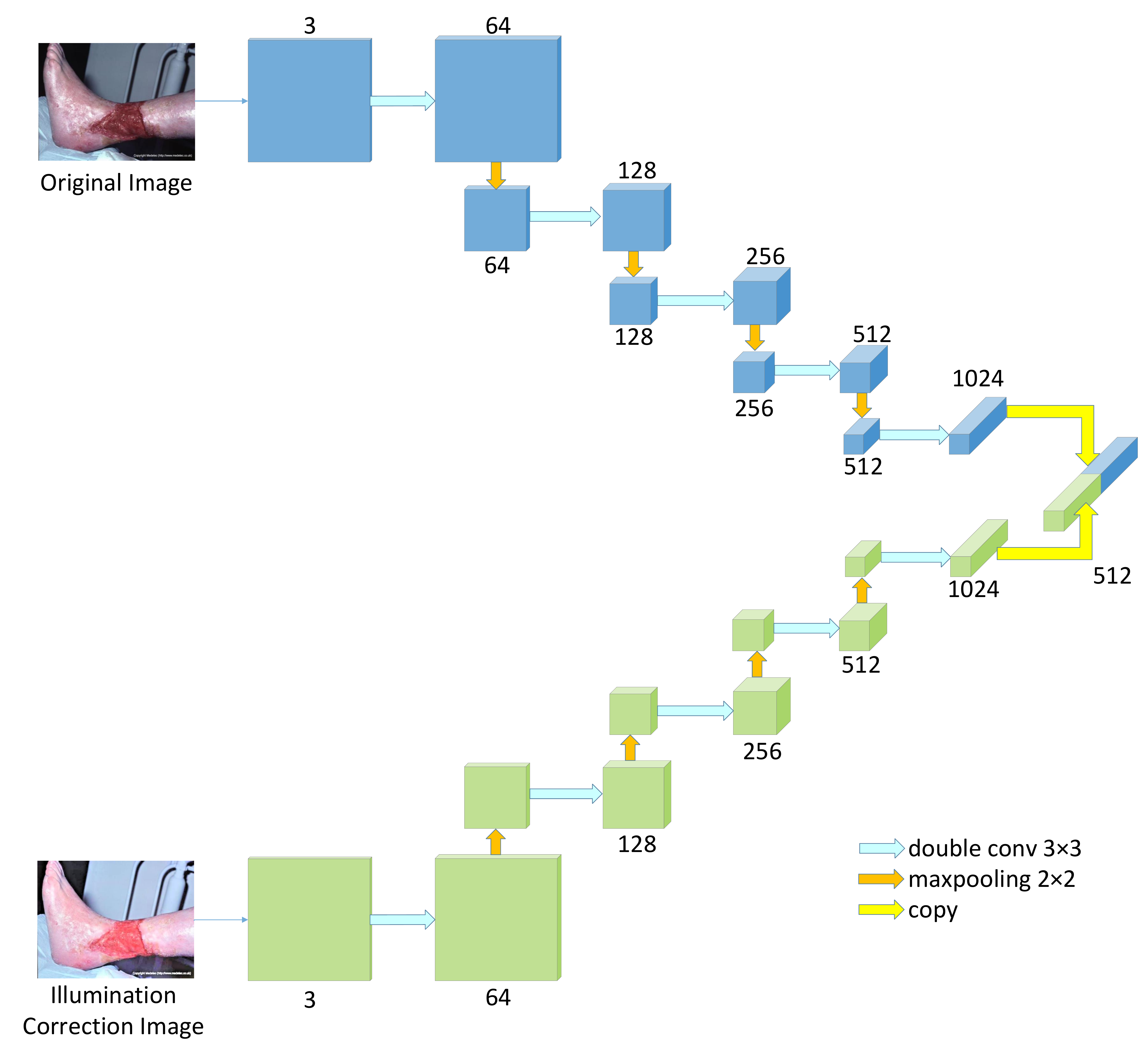}
	\caption{Feature Extraction from Dual-views.}
	\label{fig4}
\end{figure}

These images, which are preprocessed by the traditional algorithm, can be used as the input for one branch, so that the features from the illumination-corrected image can be learned using the network. Original images are also inputted to the network, which is in the same position as the preprocessed image. By this way, the information of the original image is retained, so that those images whose accuracy has been reduced after pre-processing also maintain the accuracy of the original image. In general, the segmentation accuracy is further improved, and the gains brought by pre-processing are also completely realized.

Both the original image and the preprocessed image are typical RGB 3-channel images.
The first step is to use the contraction path to extract features. The basic modules of the contraction path are two sets of 3$\times$3 convolutions and a 2$\times$2 Max-pooling layer. After operations four times, feature maps with 1024 channels are obtained finally, and its scale is 1/16 of the original image.

\begin{figure}[htbp]
	\centering
	\includegraphics[width=8cm]{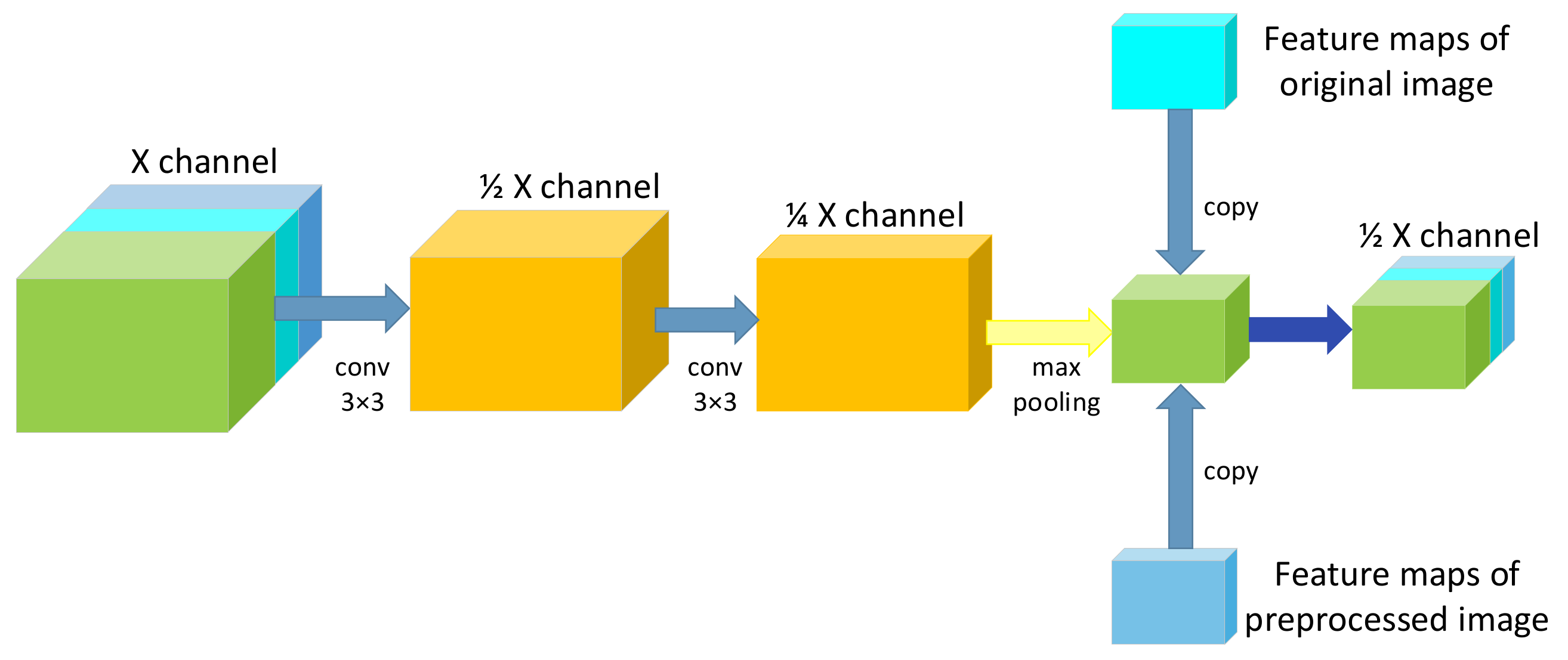}
	\caption{Fundamental Model of Contracting Path.}
	\label{fig7}
\end{figure}

The feature maps of the original image and the preprocessed image are spliced together to form a feature map of 2048 channels. This is the first step of the feature fusion. The next step is to restore the feature map to the size of the original image through the expansion path. The expansion path includes convolutional layer, up-sampling layer, and Skip-Connection. The up-sampling layer is to restore the feature map to the size of the original image. The convolutional layer can reduce the number of channels while training the convolution parameters. Skip-Connection combines the high-resolution feature maps in the contraction path with the same-scale feature maps in the expansion path by copying. This method has two advantages. (1) It combines high-resolution feature maps, strengthens the information of the border and finally improves the accuracy of the border in the final segmentation result. (2) The fusions of original image and preprocessed image features are further strengthened.

\subsection{Overall Segmentation Framework}
\begin{figure*}[]
	\centering
	\includegraphics[width=16cm]{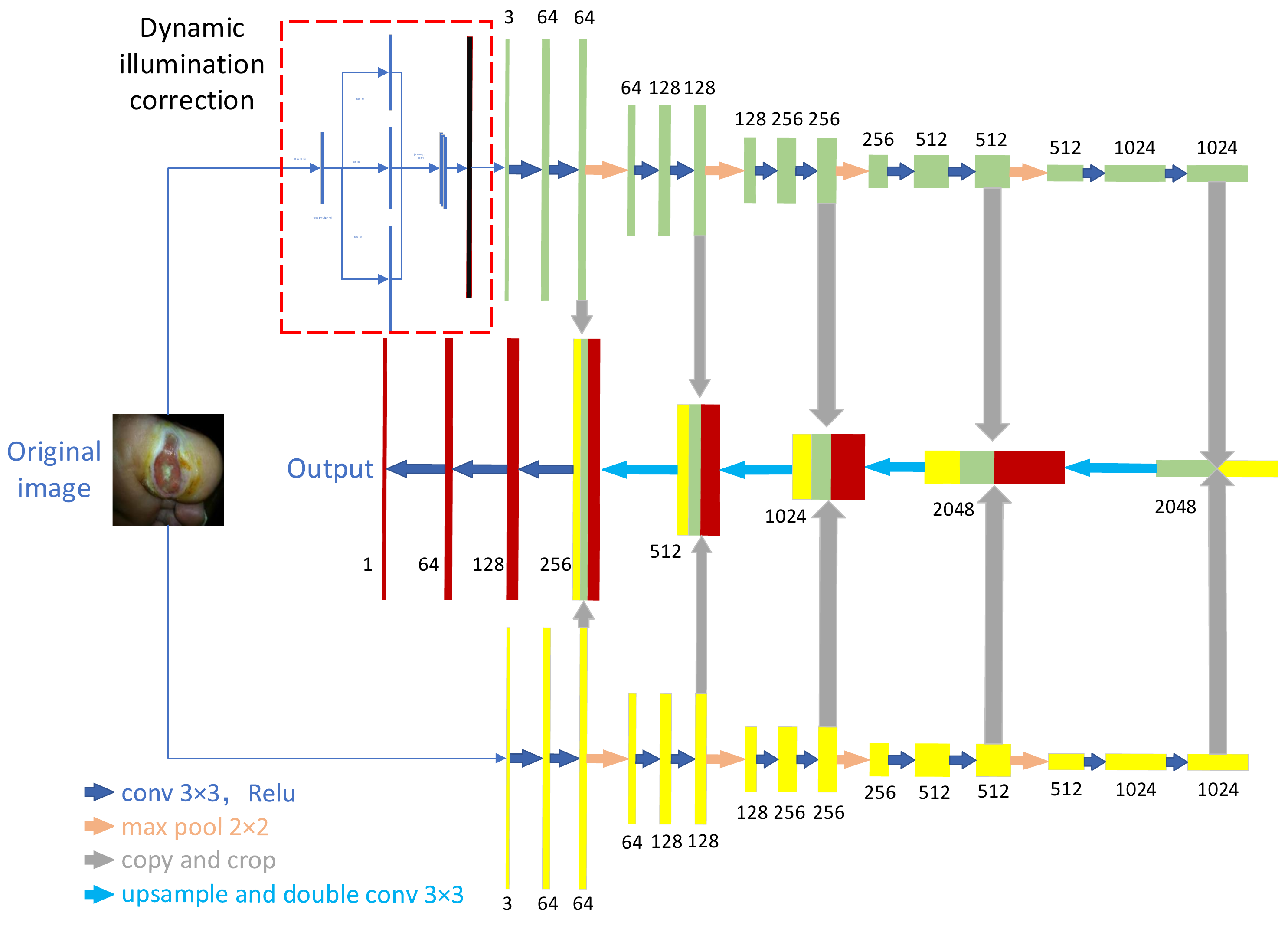}
	\caption{Overall framework of proposed method, which consists of the dynamic illumination correction module and the dual-view fusion part.}
	\label{fig5}
\end{figure*}

With previous modifications, we denote the whole network as ``D-Unet'', whose structure is demonstrated in Fig. \ref{fig5}. This network combines the first two networks to implement the dynamic illumination correction and retain the characteristic information of the original image at the same time. This network solves the problem that the traditional pre-processing algorithm cannot be adjusted according to training, and also addresses the problem that the pre-processing algorithm will reduce the accuracy of some images.

The network can be divided into two branches, a contraction path and an expansion path. The contraction path is used to extract original image features and pre-process image features. The basic module is composed of two sets of 3 $\times$ 3 convolution kernels plus a max-pooling. As the size of the feature map is reduced by half, the number of channels is doubled. First, for a typical RGB three-channel image, two 64-channel 3$\times$ 3 convolution kernel groups are used, and then the feature map size is reduced to one-half through max-pooling. After four times operation like that, feature maps of 512 channels are obtained, and the size of which is 1/16 of the original image. Meantime, the original image is processed with the dynamic illumination correction module to obtain the enhanced image, and then extract the feature by the same network.
In the same way, 512-channel, 1/16 feature maps are obtained. Later, The feature maps obtained from the two branches are merged to obtain a feature map with 1024 channels and a size of 1/16. After the above steps, a feature map group was received that combines the original image and preprocessed image information at the bottom layer.

The expansion path is used to restore the feature map to the size of the original image. The basic module is an up-sampling and two sets of three-by-three convolutions. After up-sampling, the size of the feature map is doubled. At the same time, the high-resolution feature map in the contraction path is extracted. The number of its channels will decrease exponentially. The final output channel number is 1, and the mask is the same size as the original image.

\section{Experiments}

\subsection{Datasets and Experimental Setup}
\begin{figure}[]
	\centering
	\includegraphics[width=8cm]{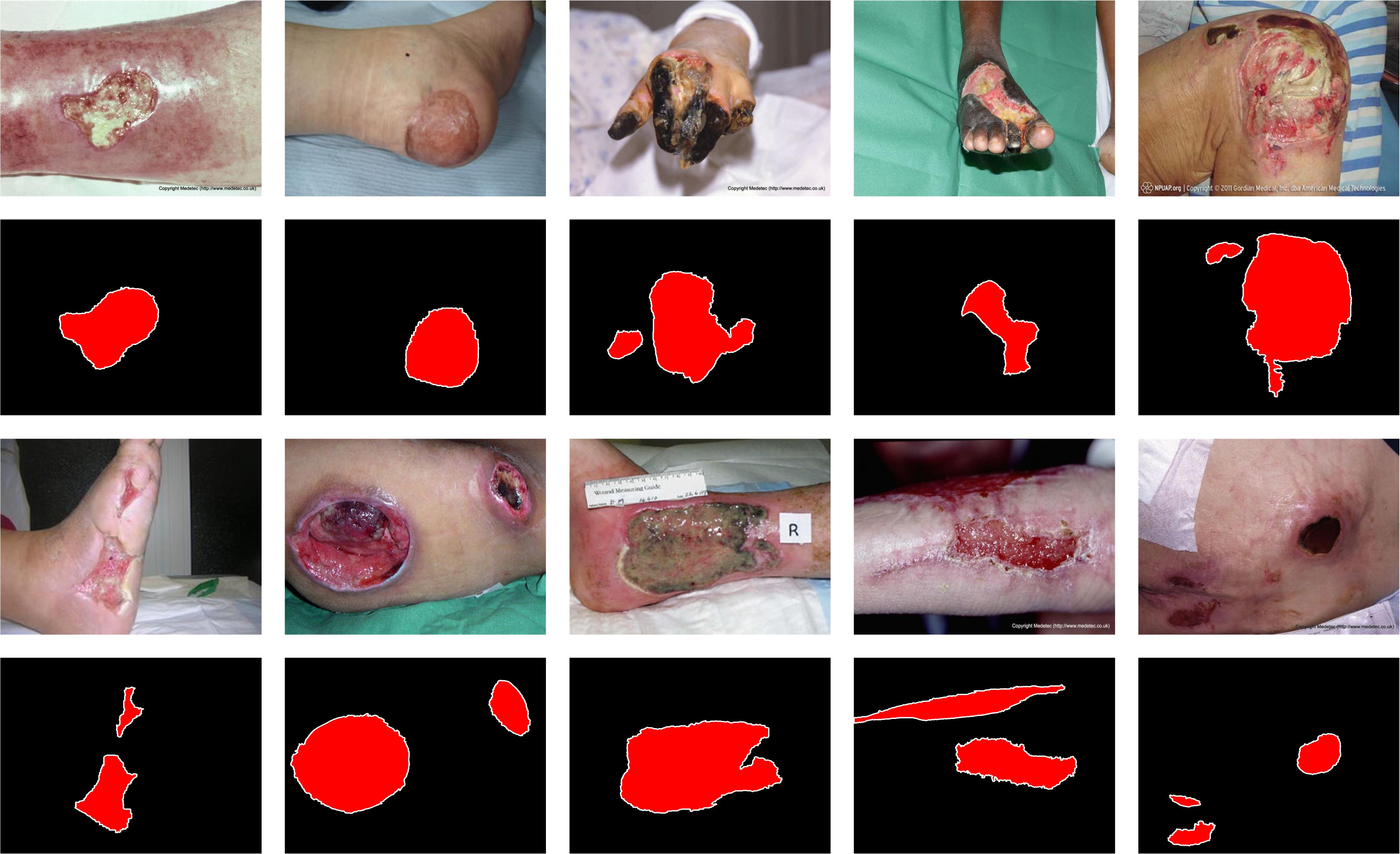}
	\caption{Sample images and masks in the dataset.}
	\label{fig6}
\end{figure}

We used the wound image dataset in \cite{wang2015a,li2018a}, which consists of 950 wound images. This data set contains some images from the Medtec Image Database, a public data set on wounds, as well as hundreds of images obtained in cooperation with hospitals, and the rest comes from the Internet. There are various types of wounds in this dataset, and the locations are wide. Most of them are diabetic feet, ulcers, burns, and many other types of wounds. The data set presents a rich and diverse form of wound expression. All images have been adjusted to 600$\times$800 pixels by cropping, rotating and other operations. Figure \ref{fig6} shows a partial image of the dataset and its corresponding mask. 

U-Net is a very representative network in semantic segmentation, and it is widely used in the field of medical image segmentation and has achieved good results. Most importantly, the Skip-connection in U-Net has great convenience for feature fusion. Therefore, U-Net was selected as the basic network of our proposed network, and it was used as the baseline to evaluate the performance of the network we propose. In this article, Baseline means the U-Net training with the original images and test by original images. The details of experiments are given as Table \ref{tab1}.

Our computing platform is 3.4GHz, 64GB RAM, and NVIDIA Titan XP. The batch size is set to 10, the optimization algorithm is RMSprop, and the loss function is CrossEntropyLoss. The learning rate is set to 0.0001.The data set was divided into two parts: training set and test set. The training set has 827 images and the remaining 122 images are used for testing. In the actual training process, 1/10 of the pictures, that is, 82 pictures, are used to verify the accuracy of the data, and the remaining 746 pictures are used to learn parameters during training. The PyTorch \footnote{https://pytorch.org/} was adopted to implement the network and we used NVIDIA Titan XP for training. The total training time was 12 hours in our experiments, while the inference can be real-time.

\begin{table}[]
	\centering
	\caption{Compare with baseline}
	\begin{threeparttable}
	\begin{tabular}{ccccccc}
		\hline
		Experiments & Network  & Train & Test & mIoU & maxIoU   \\ \hline
		group1(Baseline)      & U-Net     & O            & O        & 69.2 & 71.3     \\ 
		group2      & U-Net     & O            & TP       & 68.9 & 70.6     \\ 
		group3      & U-Net     & TP           & TP       & 69.1 & 71.2     \\ 
		group4      & DICU & O            & O        & 69.5 & 71.6     \\ 
		group5      & D-UNet   & O            & O        & \textbf{72.1} & \textbf{72.9} \\ \hline
	\end{tabular}
	\begin{tablenotes}
		\item[1] O:Original Image; TP:Traditional Pre-processing.
	\end{tablenotes}
	\end{threeparttable}
	\label{tab1}
\end{table}

As shown in the Table, IoU (intersection over union) was selected as the evaluation metrics, as it is widely used in the field of semantic segmentation. IoU can be used to evaluate the overlap between the Ground Truth and Prediction region. We can divide the pixels into True Positive (TP), False Positive (FP) or False Negative (FN), where TP stands for the area of intersection between ground truth and segmentation mask, FP stands for The predicted area outside the Ground Truth, FN stands for number of pixels in the Ground Truth area that the model failed to predict. IoU can be formulated as the following Equation \ref{eq5}. In addition, we employ \emph{Exc} (Exceed) as another metric to verify the performance of the network, which represents each image in the test set and predicts more accurately using the experimental network than using the baseline. Both of the two aforementioned metrics are used to evaluate the improvement of each image in the test set.

\begin{equation}
IoU=\frac{TP}{TP + FP + FN} \label{eq5}
\end{equation}

\begin{equation}
	Exc=\frac{Num_{inc}}{Num_{all}} \label{eq3}
\end{equation}
\noindent where $Num_{inc}$ represents the number of images whose accuracy rate is higher than that of the baseline in the control experiment. $Num_{all}$ represents the number of images in the test set. Divide the amount of improvement by the total number of test sets to get a percentage. The overall performance of the network was checked firstly.  It can be seen from the table that our network, mIoU and maxIoU have been improved significantly. This also preliminary illustrates the effectiveness of our network.

\subsection{Assessment of dynamic illumination correction}
In this section, the network for the first problem was evaluated in this article, that is, the traditional preprocessing algorithm is fixed. We designed a dynamic preprocessing module to solve this problem. We will evaluate the dynamic preprocessing algorithm from the theoretical analysis and experimental verification.

Firstly, we will analyze the performance of the dynamic preprocessing modules theoretically. We convert Gaussian filtering, which is widely using in traditional preprocessing algorithms into a convolution layers. If the training is sufficient, the result of the dynamic preprocessing algorithm must be better than the traditional fixed preprocessing algorithm. If the parameters of the convolutional layer are consistent with the Gaussian filtering, the result is the same as the traditional preprocessing algorithm In the training process, we initialize the parameters of the convolutional layer to Gaussian filtering, and remove the BatchNorm layer after the general convolutional layer. This can be closer to the optimal solution of the algorithm. In addition to theoretical analysis, a large amount of experiments have been designed and completed to evaluate dynamic preprocessing algorithms.

Table \ref{tab2} showed that the IoU of the model which trained with traditional preprocessing algorithms is even lower than the baseline. The experiments of \emph{Group3} and \emph{Group4} use dynamic preprocessing module, and it can be discovered that both \emph{mIoU} and \emph{maxIoU} is higher than Baseline.

Images are preprocessed while training in \emph{DICU} and \emph{D-Unet} by dynamic preprocessing module. The dynamic preprocessing module corrected the illumination of an image and bring out a new image. We use these image preprocessed by the dynamic illumination correction algorithm as a new dataset.
In order to evaluate the dynamic preprocessing module, another table is made of training by the new dataset.

\begin{table}[h]
	\centering
	\caption{Experimental settings used in our studies.}
	\begin{threeparttable}
	\resizebox{0.5\textwidth}{!}{
	\begin{tabular}{cccccc}
		\hline
		Experiments      & Network  & Train & Test & mIoU          & maxIoU        \\ \hline
		group1(baseline) & Unet     & O            & O        & 69.2          & 71.3          \\ 
		group4           & DICU & O            & O        & 69.5          & 71.6          \\ 
		group5           & D-Unet   & O            & O        & \textbf{72.1} & \textbf{72.9} \\ 
		group6           & Unet     & DPN1         & DPN1     & 69.5          & 71.6          \\ 
		group7           & Unet     & DPDU         & DPDU     & 69.4          & 71.4          \\ \hline
	\end{tabular}}
	\begin{tablenotes}
		\item
		 For simplicity, we denote the setting using the abbreviations. 
		 
		 O:Original Image; TP:Traditional Preprocessing;
		
		DPN1: Dynamic preprocessed images of DICU; 
		
		DPDU: Dynamic preprocessed images of D-Unet.
	\end{tablenotes}
	\end{threeparttable}
	\label{tab2}
\end{table}

\begin{figure}[!t]
	\centering
	\includegraphics[width=8cm]{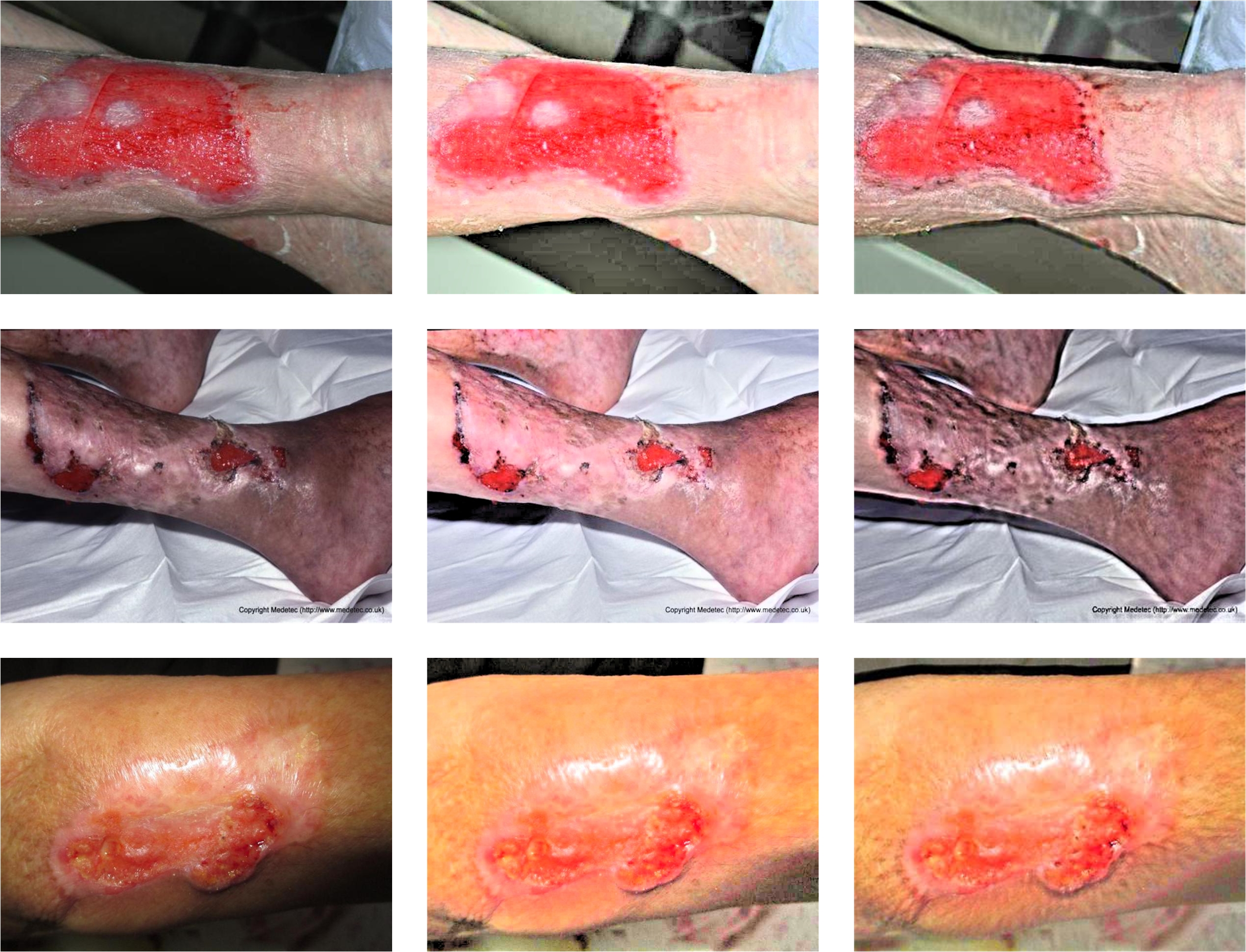}
	\caption{Differences between traditional Preprocess and our DICU method. Left: Original Image; Middle:Traditional Preprocess; Right: Dynamic Preprocess.}
	\label{fig8}
\end{figure}

The dynamically preprocessed image was extracted from the network. And it was compared with the original image and the traditional preprocessed image, as shown in figure \ref{fig8}. The main reason why traditional preprocessing causes the accuracy of partial image segmentation to decrease is that traditional preprocessing will make the brightness of some images too high, which affects the accuracy of recognition. After the adjustment of dynamic preprocessing, the brightness of the image obtained is significantly lower than that of traditional preprocessing. Thereby increasing the accuracy of segmentation. To have an intuitive understanding of the processing part, we provided certain sample results in Figure \ref{fig8}. The bottom row highlights our method performs better by visual inspection. 

As can be seen from the the above experiment and analysis, the dynamic preprocessing algorithm has shown obvious advantages compared with the traditional preprocessing algorithm.

\subsection{Assessment of dual-view fusion}
Another major problem to be solved in this paper is that the preprocessing algorithm may cause the accuracy of partial image segmentation to decrease. This problem was addressed by preserving the features of the original image in the network. It can be revealed from Table \ref{tab3} that the accuracy of most images of the network that uses the dual-view U-Net structure to extract features and merge in the later stage exceeds the accuracy of the original image. Especially in the final network, this proportion reached $97\%$. In our case, the inputs retain the features from the original images and the illumination-corrected images, therefore, the features can be fusion-ed using our framework to improve the segmentation performance. Intuitively, we expect the fusion can capture multi-scale features under different imaging setting. 

\begin{table}[]
	\centering
	\caption{The gain of illumination correction on accuracy}
	\begin{threeparttable}
	\resizebox{0.5\textwidth}{!}{
	\begin{tabular}{ccccccc}
		\hline
		Experiments      & Network  & Train & Test & mIoU          & maxIoU        & Exceed        \\ \hline
		group1(baseline) & U-Net     & O            & O        & 69.2          & 71.3          & -             \\ 
		group2           & U-Net     & O            & TP       & 68.9          & 70.6          & 33.7          \\ 
		group3           & U-Net     & TP           & TP       & 69.1          & 70.2          & 43.7          \\ 
		group8           & DVSFN & O\&TP        & O\&TP    & 69.7          & 71.5          & 79.6          \\ 
		group5           & D-UNet   & O            & O        & \textbf{72.1} & \textbf{72.9} & \textbf{86.4} \\ \hline
	\end{tabular}}
	\begin{tablenotes}
	\item
O:Original Image; TP:Traditional Prepossessing.
\end{tablenotes}
\end{threeparttable}
	\label{tab3}
\end{table}

\begin{table}[!htbp]
	\centering
	\caption{Quantitative Comparison with state-of-the-art models.}
    \resizebox{0.5\textwidth}{!}{
	\begin{tabular}{cccc}
		\hline
		Methods & Network         & Preprocessing              & IoU           \\ \hline
		Baseline     & Unet            & -                       & 75.5         \\ 
		Wang et al.  & Encoder-Decoder & -                       & 73.4          \\ 
		Li et al.    & MobileNet            & Skin detect in Cr channel                        & 82.3         \\ 
		Lu et al.    & Encoder-Decoder & Illumination Correction & 81.4          \\ 
		Ours         & D-Unet          & Dynamic Preprocess      & \textbf{84.6} \\ \hline
	\end{tabular}}
	\label{tab4}
\end{table}

\subsection{Comparison with state-of-the-art models}
To further demonstrate the performance of our proposed method, we also conducted a thorough evaluation on the wound segmentation task, with quantitative comparison to state-of-the-art models. The results of the comparison are shown in Table \ref{tab4}. Our model achieved the best results. As can be seen from the table, our method can achieve an IoU of $84.6\%$ using the datasets. We attribute the performance improvement to the dynamic illumination correction module and the dual-view fusion framework.

\section{Conclusion}
In this paper, a novel framework for the wound image segmentation was proposed, in which joints learn the preprocessing parameters from the raw and segmentation task. By incorporating the correction module into the segmentation network, our method can remove the unevenness of illumination and image enhancement is added. Moreover, a dual-view fusion strategy is proposed to extract the features from both the original image and the enhanced image. Accompanying multiple Skip-Connection to merge the previous features, extensive experiments have been conducted on the real-world wound dataset, and our proposed method provided encouraging results which indicate the benefits of jointly training the preprocessing module and the segmentation network. Our method achieved state-of-the-art performance, with comparison with existing methods.


%

\section*{Acknowledgment}
This work was supported in part by the Advanced Research Project of China (31511010203).




\bibliographystyle{IEEEtran}
\bibliography{example}

\end{document}